# Synthesis of $Fe_3O_4$ nanoparticles and their magnetic properties

Yan Wei[a], Bing Han[b], Xiaoyang Hu[a], Yuanhua Lin[c],
Xinzhi Wang[d], Xuliang Deng[a],[*]

[a].Department of Geriatric Dentistry, School and Hospital of Stomatology, Peking University, Beijing, 100081, China.
[b]. Department of orthodontics , School and Hospital of Stomatology, Peking University, Beijing, 100081, China.
[c].State Key Laboratory of New Ceramics and Fine Processing, Department of Materials Science and Engineering, Tsinghua University, Beijing, 100084, China.
d.Department of Prosthodontics , School and Hospital of Stomatology, Peking University, Beijing, 100081,. China )

**Abstract**

$Fe_3O_4$ magnetic nanoparticles (MNPs) were synthesized by a co-precipitation method using sodium citrate and oleic acid as modifiers. Phase composition and microstructure analysis indicate that the sodium citrate and oleic acid have been successfully grafted onto the surface of $Fe_3O_4$ MNPs. The magnetic behaviors reveal that the modification can decrease the saturation magnetization of $Fe_3O_4$ MNPs due to the surface effect. $Fe_3O_4$ MNPs modified by sodiumcitrate and oleic acid show excellent dispersion capability, which should be ascribed to the great reduction of high surface energy and dipolar attraction of the nanoparticles.

## 1. Introduction

Compared to atomic or bulky counterparts, nano-sized materials owe superior physical and chemical properties due to their mesoscopic effect, small object effect, quantum size effect and surface effect. Recently, $Fe_3O_4$ MNPs have been intensively investigated because of their superparamagnety, high coercivity and low Curie temperature[1–4]. In addition to these characterss, $Fe_3O_4$ MNPs are also non-toxic and biocompatible. Therefore, $Fe_3O_4$ MNPs have brought out some new kinds of biomedical applications such as dynamic sealing[5], biosensors[6], contrasting agent in magnetic resonance (MR) imaging[7], localizer in therapeutic hyperthermia[8] and magnetic targeted-drug delivery system[9], etc.

---

[*] Corresponding author. X.L .Deng Tel.: +86-10-62173403; fax: +86-10-62179977-2584.
*E-mail address:* kqdengxuliang@bjmu.edu.cn.
Co-Corresponding author:Yuanhua Lin Tel.: +86-10-62773741; fax: +86-10- 62771160.
*E-mail address:* linyh@mail.tsinghua.edu.cn.

It is well known that it is very important to ensure the narrow size distribution, good dispersion and high magnetic response of $Fe_3O_4$ MNPs in tissue fluid for applications. However, magnetic attractive forces combined with inherently large surface energies (>100 dyn/cm) make them easy for the aggregation $Fe_3O_4$ MNPs in fluids. Therefore, lots of synthesized polymers[10-14] (e.g., poly (vinyl alcohol) phosphate, polyethylene glycol, polyamides, polyglycidyl methacrylate, poly(acrylic acid), chitosan (CS) and o-carboxymethylchitosan) were employed as coating agent in order to modify the surface of iron oxide particles. Although the polymeric coatings can reduce the aggregation of MNPs, they also increase the overall size of the particles and thus limit the expression of magnetic property, tissue distribution, metabolic clearance and penetration ability into interstitial spaces. So it is very important to develop an efficient surface-modification method for preparing $Fe_3O_4$ MNPs with narrow size distribution and excellent dispersion in aqueous or inaqueous solution using small moleculer compounds.

In this work, we developed an efficient modification method using sodiumcitrate and oleic acid for the synthesis of $Fe_3O_4$ MNPs with narrow size distribution and excellent dispersion in fluids. $Fe_3O_4$ MNPs were synthesized by a co-precipitation method at different temperatures, and modified with odiumcitrate and oleic acid respectively. The effect of temperature and modifiers on the crystal structure, morphology, dispersion and size distribution, and magnetic properties of $Fe_3O_4$ MNPs were investigated in detail.

## 2. Experimental

The reagents of analytic grade ($FeCl_3 \cdot 6H_2O$, $FeCl_2 \cdot 4H_2O$, NaOH and $C_2H_6O$) were used as raw materials. Chemical grade sodiumcitrate ($Na_3C_6H_5O_7 \cdot 2H_2O$) and oleic acid ($C_{17}H_{33}COOH$) were used as modifiers. Four samples were prepared depending on their synthesis conditions (shown in Table 1). Firstly, $FeCl_3 \cdot 6H_2O$ and $FeCl_2 \cdot 4H_2O$ with molar proportion of 1:2 were dissolved in ethanol or deionized water maintained at different temperatures (Table 1), and then NaOH solution (3 mol·$L^{-1}$) was added into the above solution using a peristaltic pump under constant magnetic stirring for 30 min, and the final pH was 10. Afterwards, the sodiumcitrate and oleic acid were respectively added into the suspensions to modify the obtained $Fe_3O_4$ MNPs for 12h. The substance obtained were aged and digested at maintained temperature for 30 min and cooled at room temperature. The resulted particles were magnetically separated and washed repeatedly with deionized water and ethanol until pH was 7. The products were then dried at 60 °C in vacuum for 6 h for further characterizations.

**Table 1.** Preparation conditions for the $Fe_3O_4$ MNPs

| Samples | Preparation conditions | | | |
|---|---|---|---|---|
| | Temperature (°C) | Sodiumcitrate Concentration (mol $L^{-1}$) | Oleic Acid Concentration (ml $L^{-1}$) | Total Fe Concentration (mol $L^{-1}$) |
| (a) | 40 | / | / | 0.15 |
| (b) | 80 | / | / | 0.15 |
| (c) | 80 | 0.43 | / | 0.86 |
| (d) | 80 | / | 0.86 | 0.86 |

The crystal structure of as-prepared samples was analyzed by X-ray diffraction (XRD) with a Rigaku D/Max-C model diffractometer using Fe target. The molecular structure of $Fe_3O_4$ MNPs was characterized by a Perkin-Elmer Paragon 1000 Fourier transform spectrometer (FT-IR) at room temperature (25 °C). The magnetic property of $Fe_3O_4$ MNPs was measured using a vibrating sample magnetometer (VSM, LakeShore 7307). The morphology of the magnetite nanoparticles were determined using transmission electron microscopy (TEM, Hitachi H-600-II, Japan).

## 3. Results and discussion

XRD measurement was used to identify the crystalline structure of the products. As shown in Fig. 1, the XRD peaks can match well with the characteristic peaks of inverse cubic spinel structure (JCPDS 19-0629), which indicate that the crystalline structure of $Fe_3O_4$ MNPs can be remained after the surface modification with sodium citrateand the oleic acid. The average crystallite size d calculated using the Debye–Scherrer equation $d = K\lambda/(\beta\cos\theta)$ are about 12.6 nm (a), 13.4 nm (b), 14.2 nm (c) and 13.8 nm (d), respectively.

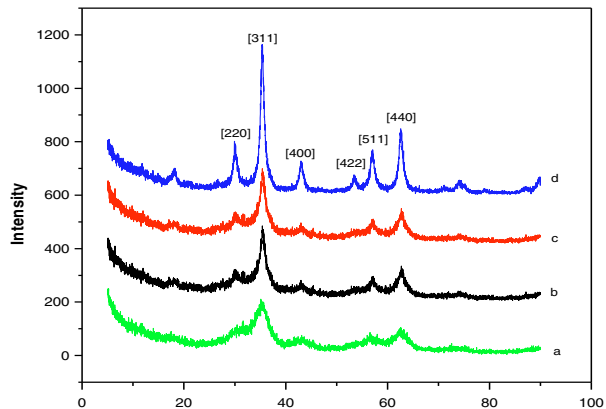

**Fig. 1.** XRD patterns of as-prepared $Fe_3O_4$ MNPs.

Figure 2 shows the FT-IR spectra of $Fe_3O_4$ MNPs of all samples. It can be seen that the characteristic absorption of Fe-O bond is at 580 $cm^{-1}$ and 634 $cm^{-1}$, while that of -OH bond is at 3398 $cm^{-1}$. The similar characteristic peaks can be found in Fig. 2b. Compared with Fig. 2b, we can see some new absorption peaks in Fig. 2c. The absorptions at 1393 $cm^{-1}$ and 1587 $cm^{-1}$ are characteristic peaks of the COO-Fe bond, which may be due to the reaction of hydroxide radical groups on the surface of $Fe_3O_4$ with carboxylate anion of sodium citrate[15]. The peaks at 2855 $cm^{-1}$ and 2924 $cm^{-1}$ are from the vibration of in long alkyl chain $-CH_2$ and $-CH_3$. Furthermore, the characteristic peak of -OH bond at 3378 $cm^{-1}$ is obviously enhanced. These peaks reveal that sodium citrate has been successfully grafted onto the surface of $Fe_3O_4$ MNPs as well. As compared with Fig. 2b, we also find some new absorption peaks in Fig. 2d. These peaks at 1384 $cm^{-1}$ and 1412 $cm^{-1}$ are attributed to the vibration of the double covalent bond in -CH=CH-, and the peaks at 2853 $cm^{-1}$ and 2924 $cm^{-1}$ correspond to $-CH_2$ and $-CH_3$ bond. Obviously, the characteristic absorption peaks of -OH bond at 3387 $cm^{-1}$ decreased, which indicates that oleic acid has been successfully grafted onto the surface of $Fe_3O_4$ MNPs through the chemical reaction of hydroxide radical groups on the surface of $Fe_3O_4$ with carboxylic acid groups of oleic acid. Both the peak at 598 $cm^{-1}$ in Fig. 2c and the peaks at 581 $cm^{-1}$ and 628 $cm^{-1}$ in Fig. 2d confirm the existence of Fe-O bond.

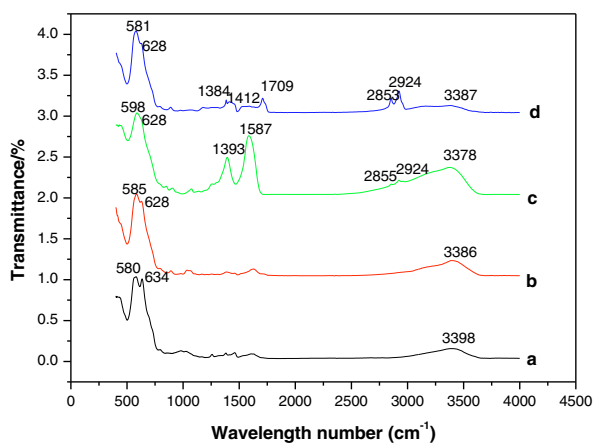

**Fig. 2.** FT-IR spectra of all the $Fe_3O_4$ samples.

M-H hysteresis curves of all samples are shown in Fig. 3. Symmetric hysteresis and saturation magnetization can be observed, and all as-prepared $Fe_3O_4$ MNPs show ferrimagnetic behaviors. This is because that the diameter of MNPs is smaller with that of critical threshold of $Fe_3O_4$ (25 nm). It can be seen that the saturation magnetization of sample a, b c and d are 50.61, 61.36, 56.05 and 55.43 $emu·g^{-1}$ respectively, which are obviously lower than that of the bulk $Fe_3O_4$ (90 $emu·g^{-1}$)[11]. This phenomenon may attribute to the small particle size effect since a noncollinear spin arrangement occurs primarily at or near the surface, which results in the reduction of magnetic moment in $Fe_3O_4$ NPs [12, 13]. The saturation magnetization of sample b is larger than that of sample a, which may be ascribed to the increase of particle size of $Fe_3O_4$ MNPs. The saturation magnetization decreases evidently when the $Fe_3O_4$ MNPs were modified with sodiumcitrate and oleic acid. This result could be attributed to the surface spin effect on $Fe_3O_4$ MNPs caused by modification, which subsequently decrease the saturation magnetization value. Nevertheless, compared to the decrease of saturation magnetization value in polymer-modification[10-14], the reduction by sodiumcitrate or oleic acid modification is significantly lower, showing the obvious advantage using small moleculer compounds as modifiers.

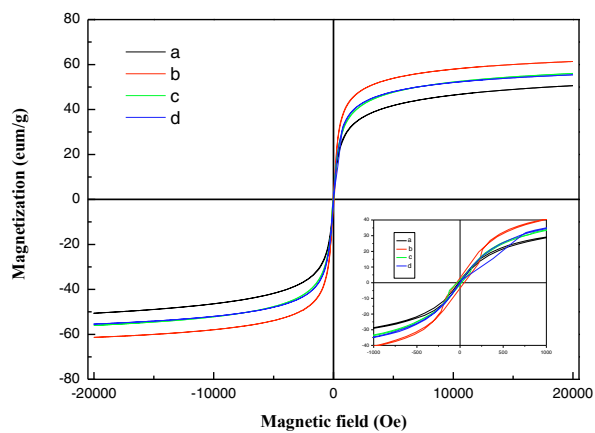

**Fig. 3.** M-H hysteresis curves of all the $Fe_3O_4$ MNPs.

The morphology and distribution of these samples were further characterized using TEM as shown in Fig. 4. All the Fe$_3$O$_4$ MNPs show homogeneously spherical shape with diameter about 12-15 nm, which is in agreement with the results of the XRD analysis. The Fe$_3$O$_4$ MNPs prepared without modification aggregate in deionized water (Fig. 4a). The sodiumcitrate and oleic acid -modified Fe$_3$O$_4$ MNPs show good dispersion capability in deionized water and oleic acid solution, which should be due to the fact that the high surface energy and dipolar attraction of the MNPs greatly reduced after modifided by sodiumcitrate and oleic acid.

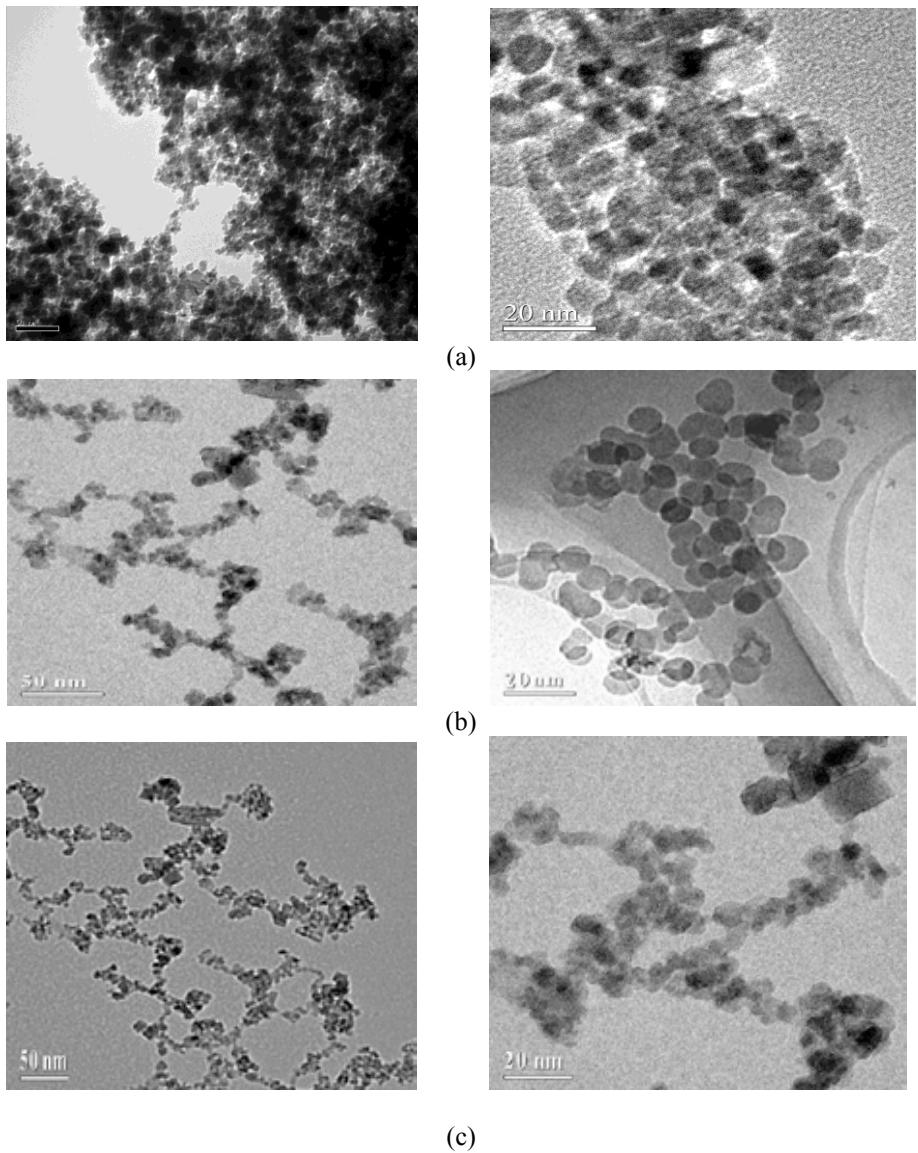

**Figure 4**. TEM images showing the Fe$_3$O$_4$ MNPs synthesized by a co-precipitation method at 80 °C. (a) Fe$_3$O$_4$ MNPs without any modifier, (b) Fe$_3$O$_4$ MNPs modified by sodiumcitrate, (c) Fe$_3$O$_4$ MNPs modified by oleic acid.

## 4. Conclusions

In summary, we have presented a simple and effective technique to prepare and modify $Fe_3O_4$ MNPs. The results indicate that the sodium citrate and oleic acid modification has little effect on the crystallization of $Fe_3O_4$ MNPs. FT-IR spectra reveals that sodium citrate and oleic acid have been successfully grafted onto the surface of $Fe_3O_4$ MNPs by chemical bond. Sodiumcitrate and oleic acid modified $Fe_3O_4$ MNPs show excellent dispersion capibility in aqueous or inaqueous solution, which make $Fe_3O_4$ MNPs as a promising biomedical material.

## Acknowledgments

The authors acknowledge financial support from National Natural Science Foundation of China (No. 51025205), the National High Technology Research and Development Program of China (2007AA03Z328766), International Science and Technology Cooperation Program (2007 DFA30690) and the Program of New Century Excellent Talents (NCET) of Universities in China. Y. Wei thanks Z. Fang for his kind help.

## References:

[1]  Kim YS, Kim YH. Application of ferro-cobalt magnetic fluid for oil sealing. *J Magn. Magn. Mater* 2003;**267**:105-110.
[2]  Raj K, Moskowitz R. A review of damping applications of ferrofluids. *Transactions on Magnetics* 2002;**16**:358-363.
[3]  Beydoun D, Amal R, Low GKC, McEvoy S. Novel photocatalyst: titania-coated magnetite. Activity and photodissolution. *J Phys Chem B* 2000;**104**:4387-4396.
[4]  McMichael RD, Shull RD, Swartzendruber LJ, Bennett LH, Watson RE. Magnetocaloric effect in superparamagnets. *J Magn Magn Mate* 1992;**111**:29-33.
[5]  Shen L, Laibinis PE, Hatton TA. Bilayer surfactant stabilized magnetic fluids: Synthesis and interactions at interfaces. *Langmuir* 1999;**15**:447-453.
[6]  Jordan A, Scholz R, Wust P, Fähling H, Felix R. Magnetic fluid hyperthermia (MFH): Cancer treatment with AC magnetic field induced excitation of biocompatible superparamagnetic nanoparticles. *J Magn Magn Mater* 1999;**201**:413-419.
[7]  Cao JQ, Wang YX, Yu JF, Xia JY, Zhang CF, Yin DZ, Hafeli UO. Preparation and radiolabeling of surface-modified magnetic nanoparticles with rhenium-188 for magnetic targeted radiotherapy. *J Magn Magn Mate* 2004;**277**:165-174.
[8]  Jordan A, Scholz R, Wust P, Schirra H, Schiestel T, Schmidt H, Felix R. Endocytosis of dextran and silan-coated magnetite nanoparticles and the effect of intracellular hyperthermia on human mammary carcinoma cells in vitro. *J Magn Magn Mate* 1999;**194**:185-196.
[9]  Li Q, Xuan YM, Wang J. Experimental investigations on transport properties of magnetic fluids. *Exp Therm Fluid Sci* 2005; **30**:109-116.
[10] Kim DK, Mikhaylova M, Zhang Y, Muhammed M. Protective Coating of Superparamagnetic Iron Oxide Nanoparticles. *Chem Mater* 2003;**15**:1617-1627.
[11] Tao K, Dou HJ, Sun K. Interfacial coprecipitation to prepare magnetite nanoparticles: Concentration and temperature dependence. *Colloids Surf* 2008;**320**:115-122.
[12] Kennedy RJ, Stampe PA. $Fe_3O_4$ films grown by laser ablation on Si (100) and GaAs (100) substrates with and without MgO buffer layers. *J Phys D: Appl Phys* 1999;**32**:16-21.
[13] Si SF, Li CH, Wang X, Yu DP, Peng Q, Li YD. Magnetic Monodisperse $Fe_3O_4$ Nanoparticles. *Cryst Growth Des* 2005;**5**:391-393.
[14] Ding Y, Hu Y, Zhang LY, Chen Y, Jiang XQ. Synthesis and Magnetic Properties of Biocompatible Hybrid Hollow Spheres. *Biomacromolecules* 2006;**7**:1766-1772.
[15] Hong RY, Zhang SZ, Di GQ, Li HZ, Zheng Y, Ding J, Wei DG. Preparation, characterization and application of $Fe_3O_4$/ZnO core/shell magnetic nanoparticles. *Mater Res Bull* 2008;**43**:2457-2468.

# Erratum

# Erratum to "Synthesis of $Fe_3O_4$ Nanoparticles and their Magnetic Properties" [Procedia Engineering. 2012, doi: 10.1016/j.proeng.2011.12.498]

In the abovementioned article, we are very sorry that the Fig.1c and Fig.4b were given incorrectly. We have tried to contact Procedia Engineering, but the journal has been discontinued. The last issue we could found online was published in 2018. Since this journal was affiliated to Elsevier publishing group, we are still trying to get contact with the Elsevier editorial office. Anyhow, we would like to firstly make erratum here in arXiv. We are sincerely sorry for the inconvenience for those interested in our research.

Fig. 1c should be corrected as follows.

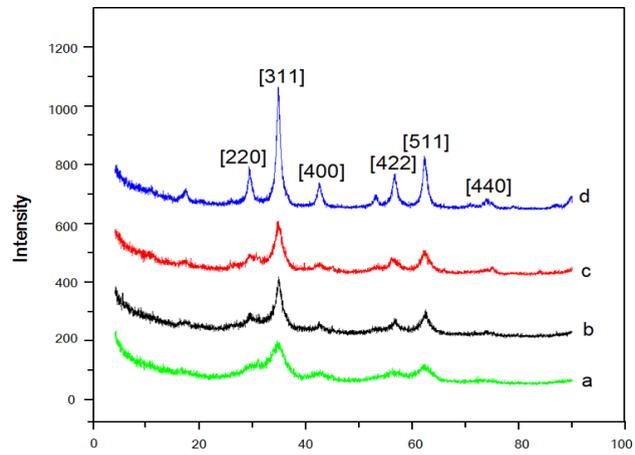

Fig. 4b should be corrected as follows.

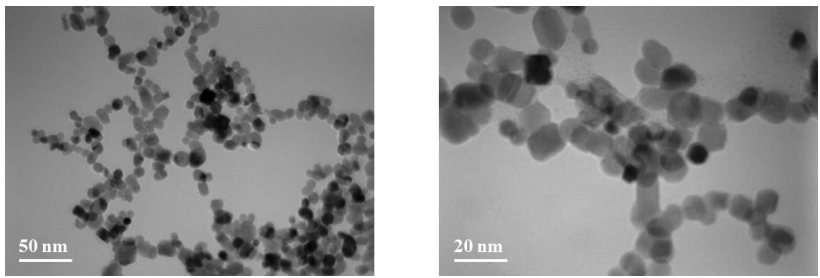